\documentclass[journal,twoside]{IEEEtran}
\usepackage{subfigure}
\usepackage{setspace}
\usepackage{amsmath}
\usepackage{amssymb}
\usepackage{amsfonts}
\usepackage{amscd}
\usepackage{mathrsfs}
\usepackage[final]{graphicx}
\usepackage{graphicx}
\usepackage{psfrag}
\usepackage{color}
\usepackage{url}
\newtheorem{theorem}{Theorem}
\newtheorem{lemma}{Lemma}
\newtheorem{definition}{Definition}
\newtheorem{proposition}{Proposition}

\newtheorem{example}{Example}

\begin{document}

\title{Co-ordinate Interleaved Distributed Space-Time Coding for Two-Antenna-Relays Networks}
\author{Harshan J~and
        B. Sundar Rajan,~\IEEEmembership{Senior Member,~IEEE}%

\thanks{This work was supported through grants to B.S.~Rajan; partly by the IISc-DRDO program on Advanced Research in Mathematical Engineering, and partly by the Council of Scientific \& Industrial Research (CSIR, India) Research Grant (22(0365)/04/EMR-II). The material in this paper was presented in parts at the IEEE Global telecommunication conference (GLOBECOM 2007), Washington D.C., USA, Nov. 26-30, 2007. Harshan J and B. Sundar Rajan are with the Department of Electrical Communication Engineering, Indian Institute of Science, Bangalore-560012, India. Email:\{harshan,bsrajan\}@ece.iisc.ernet.in.}
\thanks{Manuscript received August 08, 2007; revised November 03, 2007.}}

\markboth{IEEE Transactions on Wireless Communications ,~Vol.~xx, No.~xx, xxxx}{Harshan \MakeLowercase{and} Rajan: Co-ordinate Interleaved Distributed Space-Time Coding for Two-Antenna-Relays Networks}

\maketitle

\begin{abstract}

Distributed space time coding for wireless relay networks when the source, the destination and the relays have multiple antennas have been studied by Jing and Hassibi. In this set-up, the transmit and the receive signals at different antennas of the same relay are processed and designed independently, even though the antennas are colocated. In this paper, a wireless relay network with single antenna at the source and the destination and two antennas at each of the $R$ relays is considered. A new class of distributed space time block codes called Co-ordinate Interleaved Distributed Space-Time Codes (CIDSTC) are introduced where, in the first phase, the source transmits a $T$-length complex vector to all the relays and in the second phase, at each relay, the in-phase and quadrature component vectors of the received complex vectors at the two antennas are interleaved and processed before forwarding them to the destination. Compared to the scheme proposed by Jing-Hassibi, for $T \geq 4R$, while providing the same asymptotic diversity order of $2R$, CIDSTC scheme is shown to provide asymptotic coding gain with the cost of negligible increase in the processing complexity at the relays. However, for moderate and large values of $P$, CIDSTC scheme is shown to provide more diversity than that of the scheme proposed by Jing-Hassibi. CIDSTCs are shown to be fully diverse provided the information symbols take value from an appropriate multi-dimensional signal set.

\end{abstract}

\section{Introduction and Preliminaries}
Co-operative diversity is proved to be an efficient means of achieving spatial diversity in wireless networks without the need of multiple antennas at the individual nodes. In comparison with single user colocated multiple antenna transmission, co-operative communication is based on the relay channel model where a set of distributed antennas belonging to multiple users in the network co-operate to encode the signal transmitted from the source and forward it to the destination so that the required diversity order is achieved \cite{SEA1,SEA2,LaW,NBK}.\\
\indent In \cite{JiH1}, the idea of space-time coding devised for point to point co-located multiple antenna systems is applied for a wireless relay network with single antenna nodes and PEP (Pairwise error probability) of such a scheme was derived. It is shown that in a relay network with a single source, a single destination with $R$ single antenna relays, distributed space time coding (DSTC) achieves the diversity of a colocated multiple antenna system with $R$ transmit antennas and one receive antenna, asymptotically. \\
\indent Subsequently, in \cite{JiH2}, the idea of \cite{JiH1} is extended to relay networks where the source, the destination and the relays have multiple antennas. But, co-operation between the multiple antennas of each relay is not used, i.e., the colocatedness of the antennas is not exploited. Hence, a total of $R$ relays each with a single antenna is assumed in the network instead of a total of $R$ antennas in a smaller number of relays. With this set up, for a network with $M$ antennas at the source, $N$ antennas at the destination and a total of $R$ antennas at $R$ relays, for large values of $P$, the PEP of the network, varies with $P$ as

\begin{eqnarray*}
 \left(\frac{1}{P}\right) ^{min(M,N)R} &\mbox{ if }& M \neq N \mbox{  and  }\\
\left(\frac{(\mbox{log}_{e}^{1/M}P)}{P}\right)^{MR} &\mbox{ if }& M = N.
\end{eqnarray*}

\noindent In particular, the PEP of the scheme in \cite{JiH2} for large $P$ when specialized to $M = N = 1$ with 2$R$ antennas at relays is upper-bounded by,

\begin{equation}
\label{JingPEP}
\left[\frac{32R}{T(\rho^\prime)^{2}}\right]^{2R}\left[\frac{(\mbox{log}_{e}(P))^{2R}}{P^{2R}}\right]
\end{equation}

\noindent where $(\rho^\prime)^{2}$ is the minimum singular value of $(\textbf{S} - \textbf{S}^\prime)^{H}(\textbf{S}- \textbf{S}^\prime)$ where $\textbf{S}$ and $\textbf{S}^\prime$ are the two distinct codewords of a distributed space time block code and $P$ is the total power per channel use used by all the relays for transmitting an information vector.\\  
Following the work of \cite{JiH2}, constructions of distributed space time block codes for networks with multiple antenna nodes are presented in \cite{OgH1}, \cite{OgH2}.

We refer cooperative diversity schemes in which multiple antennas of a relay do not co-operate i.e when the transmitted vector from every antenna is function of only the received vector in that antenna, or when every relay has only one antenna as Regular Distributed Space-Time Coding (RDSTC).\\
The key idea in the proposed scheme is the notion of vector coordinate interleaving defined below:                                                                                         
\begin{definition}\label{def1} Given two complex vectors $\textbf{y}_{1}, \textbf{y}_{2} \in \mathbb{C}^{T},$ we define a Coordinate Interleaved Vector Pair of $\textbf{y}_{1}, \textbf{y}_{2},$ denoted as  $\mbox{CIVP}\left\{ \textbf{y}_1, \textbf{y}_2\right\}$ to be the pair of complex vectors $\left\{ {\textbf{y}_1^\prime, \textbf{y}_2^\prime}\right\},$ where  $\textbf{y}_1^\prime, \textbf{y}_2^\prime \in \mathbb{C}^{T},$ given by
\begin{equation*}
\textbf{y}_{1}^\prime = \mbox{Re} \,\textbf{y}_{1} + \textbf{j} \,\mbox{Im}\, \textbf{y}_{2}, ~ \textbf{y}_{2}^\prime = \mbox{Re} \,\textbf{y}_{2} + \textbf{j} \,\mbox{Im}\, \textbf{y}_{1},
\end{equation*}
or equivalently,
\begin{equation}\label{civp1}
\textbf{y}_{1}^\prime = (\textbf{y}_{1} + \textbf{y}_1^* + \textbf{y}_{2} - \textbf{y}_2^*)/2,
\end{equation}
\begin{equation}\label{civp2}
\textbf{y}_{2}^\prime = (\textbf{y}_{2} + \textbf{y}_2^* + \textbf{y}_{1} - \textbf{y}_1^*)/2.\\
\end{equation}
\end{definition}

The notion of coordinate interleaving of two complex variables has been used in \cite{KhR} to obtain single-symbol decodable STBCs with higher rate than the well known complex orthogonal designs. Definition \ref{def1} is an extension of the above technique to two complex vectors. The idea of vector co-ordinate interleaving has been used in \cite{WuB} in order to obtain better diversity results in fast fading MIMO channels.

\indent In this paper, we show that multiple antennas at the relays can be exploited to improve the performance of the network. Towards this end, a single antenna source and a single antenna destination with two antennas at each of the $R$ relays is considered. Also, the two phase protocol as in \cite{JiH2} is assumed where the first phase consists of transmission of a $T$ length complex vector from the source to all the relays (not to the destination) and the second phase consists of transmission of a $T$ length complex vector from each of the antennas of the relays to the destination, as shown in Fig.\ref{model}. The modification in the protocol we introduce is that the two received vectors at the two antennas of a relay during the first phase is coordinate interleaved as defined in Definition \ref{def1}. Then, multiplying the coordinate interleaved vector with the predecided antenna specific $T\times T$ unitary matrices, each antenna produces a $T$ length vector that is transmitted to the destination in the second phase. The collection of all such vectors, as columns of a $T\times 2R$ matrix constitutes a codeword matrix and collection of all such codeword matrices is referred as coordinate interleaved distributed space time code (CIDSTC). The contributions of this paper may be summarized as follows in more specific terms: 
\begin{itemize}
\item For $T \geq 4R$, an upper bound on the PEP of our scheme with fully diverse CIDSTC, at large values of the total power $P$ is derived.
\item For $T \geq 2R,$ the PEP of the RDSTC scheme in \cite{JiH2} with fully diverse DSTBC is upper bounded  by the expression given in \eqref{JingPEP}. Comparing this bound, with ours, for equal number of $2R$ antennas, a term $\left[ \mbox{log}_{e}(P)\right]^{R}$ appears in the numerator of the PEP expression of our scheme instead of the term $\left[ \mbox{log}_{e}(P)\right]^{2R}$. This improvement in the PEP comes just by vector co-ordinate interleaving at every relay the complexity of which is negligible.
\item It is shown that CIDSTC scheme provides asymptotic coding gain compared to the corresponding RDSTC scheme.
\item CIDSTC in variables $x_{1}, x_{2} \cdots x_{T}$ is shown not to provide full diversity if the variables $x_{1}, x_{2} \cdots x_{T}$ take values from any 2-dimensional signal set. 
\item Multi-dimensional signal sets are shown to provide full diversity for CIDSTCs whose choice depends on the design in use.
\item The number of channel uses needed in the proposed scheme is at least $4R$ where as only $2R$ is needed in an RDSTC scheme. With $T=4R$ for both the schemes, through simulation, it is shown that CIDSTC gives improved BER (Bit Error Rate) performance over that of RDSTC scheme.
\end{itemize}
\textit{Notations:} Through out the paper, boldface letters and capital boldface letters are used to represent vectors and matrices respectively. For a complex matrix \textbf{X}, the matrices $\textbf{X}^*$, $\textbf{X}^T$,  $\textbf{X}^{H}$, $\mbox{det}\left[\textbf{X}\right]$, $||\textbf{X}||_{F}^{2}$, $\mbox{Re}~\textbf{X}$ and $\mbox{Im}~\textbf{X}$ denote the conjugate, transpose, conjugate transpose, determinant, Frobenious norm,  real part and imaginary part of \textbf{X} respectively. $\textbf{I}_T$ and $\textbf{0}_T$ denotes the $ T\times T$ identity matrix and  the $T \times T$ zero matrix respectively. Absolute value of a complex number $x$, is denoted by $|x|$ and $E \left[x\right]$ is used to denote the expectation of the random variable $x.$ A circularly symmetric complex Gaussian random vector $\textbf{x}$ with mean $\mu$ and covariance matrix $\mathbf{\Gamma}$ is denoted by $\textbf{x} \sim \mathcal{CG} \left(\mu, \mathbf{\Gamma} \right)$. The set of all integers and complex numbers are denoted by $\mathbb{Z}$ and $\mathbb{C}$ respectively and $\textbf{j}$ is used to denote $\sqrt{-1}.$ Through out the paper $\mbox{log} (.)$ refers to $\mbox{log}_{e}(.)$.\\
\indent The remaining content of the paper is organized as follows: In Section \ref{sec2}, the signal model and a formal definition of CIDSTC is given along with an illustrative example.
The pairwise error probability (PEP) expression for a CIDSTC is obtained in Section \ref{sec3} using which it is shown that (i) CIDSTC scheme gives asymptotic diversity gain equal to the total number of antennas in the relays and (ii) offers asymptotic coding gain compared to the corresponding RDSTCs. Constructions of CIDSTCs along with conditions on the full diversity of CIDSTCs are provided in Section \ref{sec4}. In Section \ref{sec5}, simulation results are presented to illustrate the superiority of CIDSTC schemes. Concluding remarks and possible directions for further work constitute Section \ref{sec6}.

\section{signal model}
\label{sec2}
\begin{figure}
\centering
\includegraphics[width=2.5in]{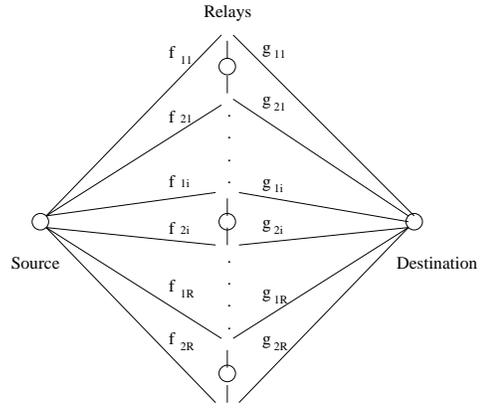}
\caption{Wireless relay network with two-antenna-relays}
\label{model}
\end{figure}

The channel from the source node to the $i$-th antenna of the $j$-th relay is denoted as $f_{ij}$ and the channel from the $i$-th antenna of the $j$-th relay to the destination node is represented by $g_{ij}$ for $i=1,2$ and $j=1,2, \cdots, R$  as shown in Fig.\ref{model}.
\noindent The following assumptions are made in our system model:

\begin{itemize}
\item All the nodes are subjected to half duplex constraint.
\item Fading coefficients  $f_{ij},g_{ij}$ are i.i.d $ \mathcal{CG} \left(0,1 \right)$ with coherence time interval, $T$.
\item All the nodes are synchronized at the symbol level.
\item Destination knows all the fading coefficients $f_{ij},g_{ij}$.
\end{itemize}


\noindent In the first phase the source transmits a $T$ length complex vector from the codebook $\mathcal{S}$ = $\left\{ \textbf{s}_{1},\, \textbf{s}_{2},\, \textbf{s}_{3},\, \cdots , \textbf{s}_{L} \right\} $ consisting of information vectors $\textbf{s}_{l} \in \mathbb{C}^{T}$ such that $E\left[\textbf{s}_{l}^{H}\textbf{s}_{l}\right]$ = 1 for all $l= 1,\cdots, L$, so that $P_{1}$ is the average transmit power used at the source every channel use. When the information vector $\textbf{s}$ is transmitted, the received vector at the $i$-th antenna of the $j$-th relay is given by
\begin{equation*}\label{rij}
\textbf{r}_{i,j} = \sqrt{P_{1}T}f_{ij}\textbf{s} + \textbf{n}_{ij},~ \textit{i} =1,2 \, \, \mathrm{and} \textit{ j} = 1,2,\cdots, R
\end{equation*} 

\noindent where $\textbf{n}_{ij} \sim \mathcal{CG} \left(0,\textbf{I}_{T} \right) $ is the additive noise vector at the $i$-th antenna of the $j$-th relay.
\noindent
In the second phase, all the relay nodes are scheduled to transmit $T$ length vectors to the destination simultaneously. In general, the transmitted signals from the different antennas of the same relay can be designed as a function of the received signals at both the antennas of the relay. We use one such technique which is very simple; every relay manufactures a $\mbox{CIVP}$ using the received vectors $\textbf{r}_{1j}$ and $\textbf{r}_{2j}$ as given in \eqref{civp1} and \eqref{civp2}, i.e., $\left\{ \textbf{r}_{1j}^\prime \textbf{r}_{2j}^\prime \right\}   = \mbox{CIVP}\left\{ \textbf{r}_{1j},\textbf{r}_{2j}\right\}.$ It is straight forward. to verify that $E\left[ (\textbf{r}_{ij}^\prime)^{H}(\textbf{r}_{ij}^\prime)\right] = \left( 1 + P_{1}\right)T.$  Each relay is equipped with a pair of fixed  $T\times T$  unitary matrices $\textbf{A}_{1j}$ and $\textbf{A}_{2j}$, one for each antenna and process the above CIVP as follows: The $1^{st}$ and the $2^{nd}$ antennas of the $j$-th relay are scheduled to transmit
\begin{equation}\label{t1j}
\textbf{t}_{1j} = \sqrt{\frac{P_{2}}{(1 + P_{1})}}\textbf{A}_{1j}\textbf{r}_{1,j} ^\prime \mbox{ and }\textbf{t}_{2j} = \sqrt{\frac{P_{2}}{(1 + P_{1})}}\textbf{A}_{2j}\textbf{r}_{2,j}^\prime
\end{equation}
\noindent respectively. The average power transmitted by each antenna of a relay per channel use is $P_{2}$. The vector received at the destination is given by
\begin{equation}\label{bfy}
\textbf{y} = \sum_{j = 1}^{R} (g_{1j}\textbf{t}_{1j} + g_{2j}\textbf{t}_{2j}) + \textbf{w}
\end{equation}

\noindent where $\textbf{w} \sim \mathcal{CG} \left(0, \textbf{I}_{T} \right)$ is the additive noise at the destination. Using \eqref{t1j} in \eqref{bfy},  \textbf{y} can be written as
\begin{equation*}\label{bfy1}
\textbf{y} = \sqrt{\frac{P_{1}P_{2}T}{(1 + P_{1})}}\textbf{S}\textbf{h} + \textbf{n}
\end{equation*}
\noindent where
\begin{itemize}
\item The additive noise, $\textbf{n}$ in the above equivalent MIMO channel is given by, 
\begin{equation*}
\label{bfN}
\textbf{n} = \sqrt{\frac{P_2}{(1 + P_1)}}\left[ \sum_{j=1}^{R} (g_{1j}\textbf{A}_{1j}\textbf{n}_{1j}^\prime + g_{2j}\textbf{A}_{2j}\textbf{n}_{2j}^\prime) \right]  + \textbf{w}
\end{equation*}
with $\left\{  \textbf{n}_{1j}^\prime , \textbf{n}_{2j}^\prime \right\} =  \mbox{CIVP}\left\{ \textbf{n}_{1j}, \textbf{n}_{2j}\right\}$. Since $\textbf{n}_{ij} \sim \mathcal{CG} \left(0, \textbf{I}_{T} \right)$, we have $\textbf{n}_{ij}^\prime \sim \mathcal{CG} \left(0, \textbf{I}_{T} \right).$\\

\item The equivalent channel  \textbf{h} is given by
\begin{equation}
\label{channel}
\textbf{h} = [\textbf{d}_{1} ~~ \textbf{d}_{2} ~~ \cdots ~~ \textbf{d}_{R} ]^T \in \mathbb{C}^{4R}\\
\end{equation}
where $\textbf{d}_{j} = [g_{1j}k_{1j} ~~~  g_{1j}k_{2j} ~~~  g_{2j}k_{1j} ~~~  -g_{2j}k_{2j}]$ for $j = 1, \cdots, R$ and $k_{1j}  = (f_{1j} + f_{2j})/2, ~ k_{2j}  = (f_{1j}^* - f_{2j}^*)/2$.\\

\item The $T \times 4R$ matrix, $$\textbf{S} = \left[  \textbf{A}_{11}\textbf{s}~~ \textbf{A}_{11}\textbf{s}^*~~ \textbf{A}_{21}\textbf{s}~~ \textbf{A}_{21}\textbf{s}^*~~  \ldots ~~ \textbf{A}_{2R}\textbf{s}~~ \textbf{A}_{2R} \textbf{s}^* \right]$$ is the equivalent codeword matrix. Henceforth, by codeword matrix will be meant only this equivalent $T \times 4R$ matrix even though the transmitted vectors from the $2R$ antennas constitute a $T \times 2R$ matrix.
\end{itemize}

\noindent The collection $\mathcal{C}$ of codeword matrices shown below when $\textbf{s}$ runs over the codebook $\mathcal{S}$,
\begin{equation}
\label{code}
\mathcal{C} = \left\{ \left[  \textbf{A}_{11}\textbf{s}~~ \textbf{A}_{11}\textbf{s}^*~~ \textbf{A}_{21}\textbf{s}~~ \textbf{A}_{21}\textbf{s}^*~~  \ldots ~~ \textbf{A}_{2R}\textbf{s}~~ \textbf{A}_{2R} \textbf{s}^* \right] \right\}
\end{equation}
\noindent will be called the Co-ordinate Interleaved Distributed Space-Time code (CIDSTC).

\begin{proposition}\label{prop} The random variables $k_{ij}$ for all $i= 1,2$ and $ j = 1,\cdots ,R.$ are independent and also $k_{ij} \sim \mathcal{CG} \left( 0, 1/2\right).$\end{proposition}
 
\begin{proof}  The proof is straight forward.
\end{proof}

\begin{example}
\label{exple1}
Consider $R=1$ and $T=4.$ Let the relay specific unitary matrices $\textbf{A}_{11}$ and $\textbf{A}_{21}$ be 
\begin{equation*}
\textbf{A}_{11} = \left[\begin{array}{cccc}
1 & 0 & 0 & 0\\
0 & 1 & 0 & 0\\
0 & 0 & 1 & 0\\
0 & 0 & 0 & 1\\
\end{array}\right];~~~ \textbf{A}_{21} = \left[\begin{array}{cccc}
0 & \textbf{j} & 0 & 0\\
1 & 0 & 0 & 0\\
0 & 0 & 0 & \textbf{j}\\
0 & 0 & 1 & 0\\
\end{array}\right].
\end{equation*}
The equivalent channel is $\textbf{h} = \left[ g_{11}k_{11} ~~  g_{11}k_{21} ~~ g_{21}k_{11} ~~ -g_{21}k_{21}\right] ^T.$
\noindent The CIDSTC is the collection of 4 $\times$ 4 matrices given by,
\begin{center}
$\mathcal{C} = \left\{ \left[  \textbf{A}_{11}\textbf{s} ~~ \textbf{A}_{11}\textbf{s}^* ~~ \textbf{A}_{21}\textbf{s} ~~ \textbf{A}_{21}\textbf{s}^*\right] ~~ :\textbf{s} \in \mathcal{S} \right\} $
\end{center}
and to be explicit, with $\textbf{s}= \left[ x_1,x_2,x_3,x_4 \right]^{T}$ where $x_1,x_2,x_3,x_4$ are complex variables which may take values from a signal set like QAM, PSK etc.

\begin{equation*}
\mathcal{C} = \left\{ \left[
\begin{array}{rrrr}
x_1 & x_1^* & \textbf{j}x_2 & \textbf{j}x_2^* \\
x_2 & x_2^* & x_1  & x_1^* \\
x_3 & x_3^* & \textbf{j}x_4 & \textbf{j}x_4^* \\
x_4 & x_4^* & x_3 &x_3^* 
\end{array}
\right]  \right\}. 
\end{equation*}
\end{example}


\section{pairwise error probability}
\label{sec3}
Since the relay specific matrices $\textbf{A}_{ij}$ are unitary, $\textbf{w}$ and $\textbf{n}_{ij}^\prime$ are independent Gaussian random variables and since $g_{ij}$ are known at the receiver, $\textbf{n}$ is a Gaussian random vector with 
\begin{center}
 $E\left[ \textbf{n}\right]$  = $ \textbf{0}_{T}$ and  $E\left[ \textbf{n}\textbf{n}^{H}\right]$   = $\left(1 + \frac{P_{2}}{(1 + P_{1})}\sum_{j = 1}^{R}(|g_{1j}|^{2} + |g_{2j}|^{2})\right)\textbf{I}_{T}$.
\end{center}
Assume that $S$ is a codeword in the CIDSTC  given in \eqref{code}. When both $k_{ij} $
and $g_{ij}$ are known, ${\textbf y}|\textbf{S}$ is also a Gaussian random vector with
\begin{center}
$E\left[{\textbf y}|\textbf{S}\right] = \sqrt{\frac{P_{1}P_{2}T}{(1 + P_{1})}}\textbf{S}\textbf{h}$ and $E\left[ \textbf{yy}^{H}| \textbf{S}\right] = (1 + \frac{P_{2}}{(1 + P_{1})}\sum_{j = 1}^{R}(|g_{1j}|^{2} + |g_{2j}|^{2}))\textbf{I}_{T}.$
\end{center}
\noindent
The maximum likelihood (ML) decoding is given by
\begin{equation}
\label{ML}
arg\, \min_{\textbf{S}} || \textbf{y} - \sqrt{\frac{P_{1}P_{2}T}{(1 + P_{1})}}\textbf{S}\textbf{h} ||_{F}^{2}.
\end{equation}
\subsection{Chernoff bound on the PEP.}
\label{subsec3}
\begin{lemma} \label{chf_bound}Assume $\textbf{S}, \textbf{S}^\prime \in \mathcal{C},$ where ${\mathcal C}$ is a CIDSTC. With the ML decoding as in \eqref{ML}, the probability of decoding to $\textbf{S}^\prime$ when  $\textbf{S}$ is transmitted given that $k_{ij},\, g_{ij}$ are known at the destination has the following Chernoff bound \cite{JiH1}:

\begin{equation}
\label{chernoff}
P\left( \textbf{S}\rightarrow \textbf{S}^\prime\right | k_{ij},\, g_{ij}) \leq \,e^{-(P^\prime \mathbf{h}^{H}\textbf{U}\mathbf{h})}
\end{equation}
where 
\begin{center}
$\textbf{U} = (\textbf{S} - \textbf{S}^\prime)^{H}(\textbf{S}- \textbf{S}^\prime)$ and 
$P^\prime = \frac{P_{1}P_{2}T}{4(1 + P_{1} + P_{2}g)}$ 
\end{center}
where $g= \sum_{j = 1}^{R}(|g_{1j}|^{2} + |g_{2j}|^{2}).$
\end{lemma}

We refer to a CIDSTC as fully diverse if $\textbf{U}$ is a full rank matrix for every codeword pair. 


\begin{lemma}\label{thm2} If $\textbf{U}$ is of full rank and the minimum singular value of $\textbf{U}$ is denoted by $\rho^2$, then the PEP in \eqref{chernoff} averaged over $k_{ij}$ satisfies
\begin{equation}
\label{theorem4eqn}
P\left( \textbf{S}\rightarrow \textbf{S}^\prime\right | g_{ij})  \leq  \prod_{j = 1}^{R} 4\left[ \frac{1}{2 + P^\prime \rho^{2}g_{j}}\right] ^{2}
\end{equation}
where $g_{j} = |g_{1j}|^{2} + |g_{2j}|^{2}.$
\end{lemma}
\begin{proof} See Appendix \ref{append2}.
\end{proof}

The power allocation problem of our model is the same as the one considered in \cite{JiH1} with $2R$ single antenna relays. As introduced in the result of Lemma \ref{chf_bound}, $P^\prime = \frac{P_{1}P_{2}T}{4(1 + P_{1} + P_{2}g)}$  where $g= \sum_{j = 1}^{R}(|g_{1j}|^{2} + |g_{2j}|^{2})$ has the gamma distribution with mean and variance being $2R$. For very large values of $R$, we can make the approximation $g \thickapprox 2R$ and hence $P^\prime = \frac{P_{1}P_{2}T}{4(1 + P_{1} + P_{2}2R)}$. Since for every channel use, the power used at the source and every antenna of a relay are $P_{1}$ and $P_{2}$ respectively, total power $P$ is $P_{1} + 2RP_{2}$. Therefore, 
\begin{equation*}
P^\prime = \frac{P_{1}P_{2}T}{4(1 + P_{1} + P_{2}2R)} \leq \frac{P^{2}T}{32R(1 + P)}.
\end{equation*}
Thus, $P^\prime$ achieves the above equality when $P_{1} = \frac{P}{2}$ and  $P_{2} = \frac{P}{4R}$.
Since we have used the approximation $g \thickapprox 2R$, the above power allocation is valid only for large values of $R$ as in \cite{JiH1}. With this optimum power allocation, when $P >>1,$ we have $\left( \mbox{from } \cite{JiH1}\right)$,
\begin{equation*}
P^\prime = \frac{P_{1}P_{2}T}{4(1 + P_{1} + P_{2}g)} \simeq \frac{PT}{8(2R +g )}.
\end{equation*}

\subsection{Derivation of Diversity order for Large R}

The upper-bound on the PEP in \eqref{theorem4eqn} needs to be averaged over $g_{1j}$'s and $g_{2j}$'s to obtain the diversity order of the CIDSTC scheme. A simple approximate derivation of the diversity order considering large number of relays in the network is presented. When $R$ is large, $g \simeq 2R$ with high probability and $P^\prime  \simeq \frac{PT}{32R}$.

\begin{theorem}\label{thm3} Assume $T \geq 4R$ and the CIDSTC is fully diverse. For large total transmit power $P$, the probability of decoding to $\textbf{S}^\prime$  when  $\textbf{S}$ is transmitted is upper bounded as

\begin{equation}
\label{theorem41}
P\left( \textbf{S}\rightarrow \textbf{S}^\prime\right) \leq \left[\frac{64R}{T\rho^{2}}\right]^{2R}\left[\frac{(\mbox{log}(P))^{R}}{P^{2R}}\right].
\end{equation}
\end{theorem}
\begin{proof} See Appendix \ref{append3}.
\end{proof}
\indent The term $\frac{(log(P))^{R}}{P^{2R}}$ in the right hand side of \eqref{theorem41} can be written as $P^{-2R(1-\frac{log(log(P))}{2log(P)})}$. Hence, the diversity of the wireless relay network with CIDSTC is 
$$2R\left(1-\frac{\mbox{log}(\mbox{log}(P))}{2\mbox{log}(P)}\right)$$ 
whereas the diversity of the scheme in \cite{JiH2} is 
$$2R\left(1-\frac{\mbox{log}(\mbox{log}(P))}{\mbox{log}(P)}\right).$$ 
Asymptotically, both the expressions $$1-\frac{\mbox{log}(\mbox{log}(P))}{\mbox{log}(P)} \mbox{ and } 1-\frac{\mbox{log}(\mbox{log}(P))}{2\mbox{log}(P)}$$ can be taken to be equal, and hence the diversity gain is approximately $2R$ in both the schemes. However, for moderate values of $P,$ the second term is larger than the first one and this difference depends on $P.$ So, our scheme performs better than the one in \cite{JiH2} by an amount that depends on $P.$


\noindent The PEP of the scheme in \cite{JiH2} for large $P$ when specialized to $M = N = 1$ with 2$R$ antennas at relays is upper-bounded by \eqref{JingPEP}.

\begin{figure}
\centering
\includegraphics[width=3in]{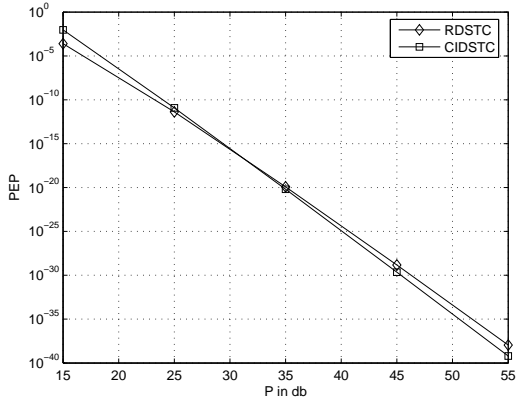}
\caption{PEP comparison : R = 5, T = 20, $\rho^\prime = 2$, $\rho = 1.5$}
\label{more}
\end{figure}

\begin{figure}
\centering
\includegraphics[width=3in]{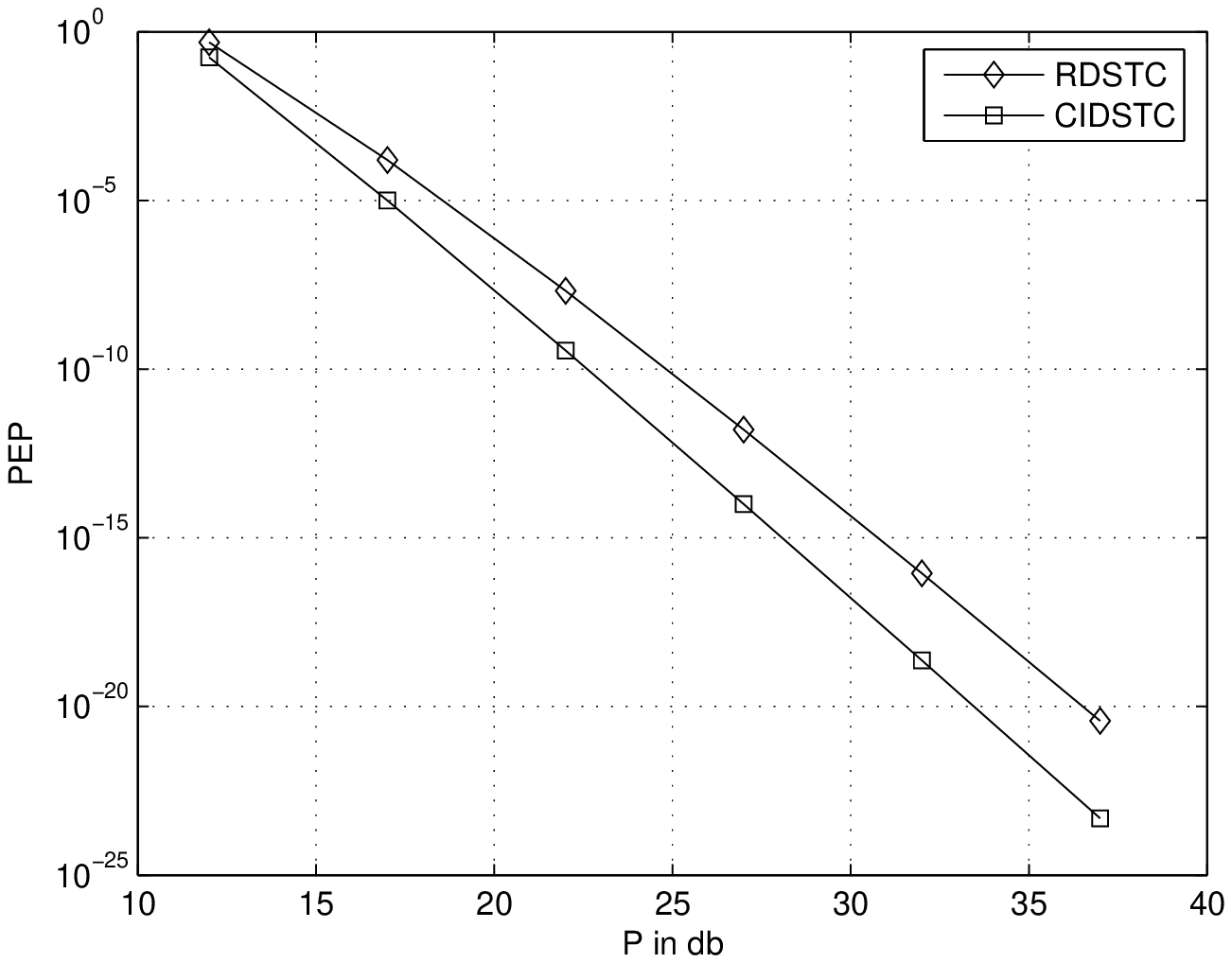}
\caption{PEP comparison : R = 5, T = 20, $\rho^\prime = 1.5$, $\rho = 2.$ }
\label{less}
\end{figure}

\begin{figure}
\centering
\includegraphics[width=3in]{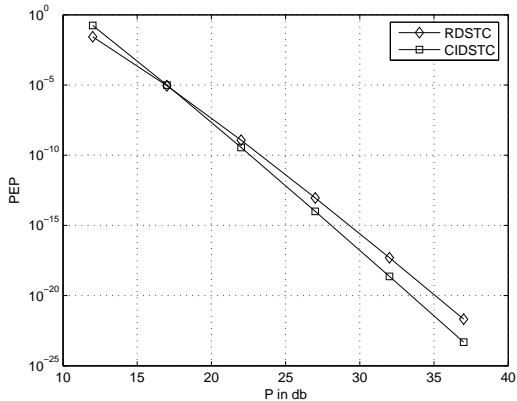}
\caption{PEP comparison : R = 5, T = 20, $\rho^\prime = 2$, $\rho = 2.$}
\label{equal}
\end{figure}

\noindent Using \eqref{JingPEP} and \eqref{theorem41}, the fractional change in PEP of CIDSTC with respect the one in \cite{JiH2} can be written as 
\begin{equation}
\label{fraction}
\left(1 - \left(\frac{2\rho^\prime}{\rho}\right)^{2R}\left(\frac{1}{\mbox{log}(P)}\right)^{R}\right).
\end{equation}
For a specified PEP, the following scenarios may occur: The total power, $P,$ required by the CIDSTC may be smaller than that of the RDSTC or vice-verse. In the former case, since we have already shown that the PEP of CIDSTC drops at a faster rate than RDSTC, the value of $P$ required to achieve a PEP below the specified PEP will be lesser for CIDSTC compared to RDSTC. 
In the event of the latter case, from \eqref{fraction} we see that, depending on the value of $\frac{\rho^\prime}{\rho}$ the corresponding value of $P$ for CIDSTC for the specified PEP may be more or less than that of the value for RDSTC. However, at large $P$, $\left(\mbox{log}(P)\right)^{R}$ dominates the above ratio and hence the expression in \eqref{fraction} increases with increase in $P$.\\
\indent Upper-bounds on the PEP in \eqref{JingPEP} and \eqref{theorem41} are plotted for $R = 5$, $T = 20$, using different values of $\rho^\prime$ and $\rho$ in Figure \ref{more}, Figure \ref{less} and Figure \ref{equal} over several values of the total power, $P$. Figures \ref{more} - \ref{equal} provide useful information on the PEP behavior of RDSTC and CIDSTC for different values of $\rho$ and $\rho^{\prime}$. In particular, these figures provide information on the power levels beyond which CIDSTC starts out performing RDSTCs and the power levels below which RDSTCs outperforms CIDSTCs. It is to be noted that the power level at which the crossover in the performance between the two schemes takes place depends on the values of $\rho$ and $\rho^{\prime}$. It is also interesting to observe that irrespective of the values of $\rho$ and $\rho^{\prime}$, there exists a sufficiently large total power $\hat{P}$ such that for $P > \hat{P}$, CIDSTC outperforms RDSTC. However, the plot shows that for all practical purposes, asymptotic coding gain provided by CIDSTC is meaningful only for the case when $\rho^{\prime} < \rho$.\\
\indent In Figures \ref{more}, \ref{less} and \ref{equal}, the upper-bounds on the PEP in \eqref{JingPEP} and \eqref{theorem41} are compared at lower values of $P$ also. Since the upper-bounds on the PEP is derived assuming a large value of $P$, the above plots may not provide actual behavior of our scheme at lower values of $P$, which corresponds to PEP in the range of $10^{-1}$ to $10^{-5}$. Plots in the above figures show that RDSTC outperforms CIDSTC at lower values of $P$ for the cases when $\rho^{\prime} > \rho$ and $\rho^{\prime} = \rho$, but we caution the reader once again to note that these plots may not provide the correct information since the derived bound is no longer valid at lower power values.

\subsection{Receiver complexity of CIDSTC}
\label{subsec3.3}
\indent From the results of Theorem \ref{thm3}, a necessary condition on the full diversity of CIDSTC is $T \geq 4R$. This implies that the number of complex variables transmitted from the source to relays in the first phase is at least twice the total number of antennas at all the relays. Therefore, CIDSTC is a design in at least 4$R$ complex variables where as a RDSTC for the same setup is a design in at least 2$R$ variables. With this, the ML decoder for CIDSTC has to decode at least 4$R$ complex variables every codeword use where as the ML decoder of RDSTC has to decode at least 2$R$ symbols every codeword use. Thus CIDSTC increases the ML decoding complexity at the receiver even though the additional complexity in performing co-ordinate interleaving of symbols at the relays is very marginal.
\section{on the full diversity of CIDSTC}\label{sec4}
In this section, we provide conditions on the signal set such that the CIDSTC in variables  $x_{1}, x_{2}, \cdots x_{T}$ is fully diverse. First, we show that CIDSTC in variables $x_{1}, x_{2}, \cdots x_{T}$ is not fully diverse if the variables take values from any 2-dimensional signal set. Towards that end, let $\textbf{X}_{\mbox{RDSTC}}$ be a fully diverse RDSTC for $2R$ relays and $T = 4R$ given by, 

\begin{equation*}
\textbf{X}_{\mbox{RDSTC}} = \left[  \textbf{A}_{1}\textbf{s}~~ \textbf{A}_{2}\textbf{s}~~  \cdots ~~ \textbf{A}_{2R}\textbf{s} \right]_{4R \times 2R}
\end{equation*}

where $\textbf{s} = \left[x_{1}, x_{2}, \cdots x_{T}\right]^{T}$ and $x_{i}$ take values from some 2-dimensional signal set. Using the above RDSTCs, we can construct a CIDSTC for $R$ relays each having two antennas by assigning the relay matrices of RDSTC to every antenna of our setup.
Therefore, CIDSTC is of the form, 

{\small
\begin{equation*}
\textbf{X}_{\mbox{CIDSTC}} = \left[  \textbf{A}_{11}\textbf{s}~  \textbf{A}_{11} \textbf{s}^*~ \textbf{A}_{21}\textbf{s}~ \textbf{A}_{21}\textbf{s}^*~ \cdots ~ \textbf{A}_{2R}\textbf{s}~ \textbf{A}_{2R}\textbf{s}^* \right]_{4R \times 4R}
\end{equation*}
}

\noindent where $\textbf{s}^{*} = \left[x_{1}^{*}, x_{2}^{*}, \cdots x_{T}^{*}\right]^{T}$. By permuting the columns, $\textbf{X}_{\mbox{CIDSTC}}$ can be written as,

\begin{equation}
\label{rep_cidstc}
\textbf{X}_{\mbox{CIDSTC}} = \left[\textbf{X}_{\mbox{RDSTC}} ~ \textbf{X}_{\mbox{RDSTC}}^{\prime}\right] 
\end{equation}

where, $\textbf{X}_{\mbox{RDSTC}}^{\prime} = \left[  \textbf{A}_{1}\textbf{s}^{*}~~ \textbf{A}_{2}\textbf{s}^{*}~~  \cdots ~~ \textbf{A}_{2R}\textbf{s}^{*} \right]_{4R \times 2R}$.\\

\noindent From \eqref{rep_cidstc}, every codeword $\textbf{S}$ of $\textbf{X}_{\mbox{CIDSTC}}$ is of the form $\textbf{S} = \left[\textbf{S}_{1} ~ \textbf{S}_{2}\right]$ where $\textbf{S}_{1} \in \textbf{X}_{\mbox{RDSTC}}$ and $\textbf{S}_{2} \in \textbf{X}_{\mbox{RDSTC}}^{\prime}$. From Section \ref{sec3}, a CIDSTC, $\mathcal{C}$ is said to be fully diverse if $\Delta \textbf{S} = \textbf{S} - \textbf{S}^{\prime}$ is full rank, for every $\textbf{S}, \textbf{S}^{\prime} \in \mathcal{C},$ such that $\textbf{S} \neq \textbf{S}^{\prime}$.\\
The difference matrix of two codewords is given by,
\begin{equation}
\label{delta}
\Delta \textbf{S} = \left[\Delta \textbf{S}_{1} ~ \Delta \textbf{S}_{2}\right]. 
\end{equation}
where $\Delta \textbf{S}_{1} = \textbf{S}_{1} - \textbf{S}_{1}^{\prime}$ and  $\textbf{S}_{1}, \textbf{S}_{1}^{\prime} \in \textbf{X}_{\mbox{RDSTC}}$ such that $\textbf{S}_{1} \neq \textbf{S}_{1}^{\prime}$. Also, $\Delta \textbf{S}_{2} = \textbf{S}_{2} - \textbf{S}_{2}^{\prime}$ and  $\textbf{S}_{2}, \textbf{S}_{2}^{\prime} \in \textbf{X}_{\mbox{RDSTC}}^{\prime}$ such that $\textbf{S}_{2} \neq \textbf{S}_{2}^{\prime}$. Since $\textbf{X}_{\mbox{RDSTC}}^{\prime}$ and $\textbf{X}_{\mbox{RDSTC}}$ are fully diverse, $\Delta \textbf{S}_{1}$ and $\Delta \textbf{S}_{2}$ are full rank.\\
\begin{equation}
\label{delta1}
\Delta \textbf{S}_{1} = \left[  \textbf{A}_{1}(\Delta \textbf{s})~~ \textbf{A}_{2}(\Delta \textbf{s})~~  \cdots ~~ \textbf{A}_{2R}(\Delta \textbf{s}) \right]_{4R \times 2R}.
\end{equation}
\begin{equation}
\label{delta2}
\Delta \textbf{S}_{2} = \left[  \textbf{A}_{1}(\Delta \textbf{s})^{*}~~ \textbf{A}_{2}(\Delta \textbf{s})^{*}~~  \cdots ~~ \textbf{A}_{2R}(\Delta \textbf{s})^{*} \right]_{4R \times 2R}
\end{equation}
where $\Delta \textbf{s} = \left[\Delta x_{1}, \Delta x_{2}, \cdots \Delta x_{4R} \right]^{T}$.\\
The following proposition shows that $\Delta \textbf{S}$ is not full rank even if $\Delta \textbf{S}_{1}$ and $\Delta \textbf{S}_{2}$ are full rank.

\begin{proposition}
\label{2_d_non_full_diversity}
If variables $x_{i}$'s take value from a 2-dimensional signal set, then CIDSTC is not fully diverse. 
\end{proposition}

\begin{proof}
Suppose complex variables $x_{i}$'s take value from any 2-dimensional signal set, then $\Delta \textbf{s}$ can possibly take values such that $\Delta x_{1} = \Delta x_{2} \cdots = \Delta x_{4R} \in \mathbb{C}$. Since $\Delta \textbf{S}_{1}, \Delta \textbf{S}_{2} \in \mathbb{C}^{4R \times 2R}$, some of the rows of $\Delta \textbf{S}_{1}$ are linearly dependent. Also, identical rows of $\Delta \textbf{S}_{2}$ will also be linearly dependent. Therefore, $\Delta \textbf{S}_{1}$ and $\Delta \textbf{S}_{2}$ together make the corresponding rows of $\Delta \textbf{S}$ linearly dependent.
\end{proof}

\begin{example}
For the CIDSTC in Example \ref{exple1}, if $\Delta x_{1} = \Delta x_{2} \cdots = \Delta x_{4} = \Delta x$, then $\Delta \textbf{S}$ is given by,

\begin{equation*}
\Delta \textbf{S} = \left[
\begin{array}{rrrr}
\Delta x & \Delta x^* & \textbf{j}\Delta x & \textbf{j}\Delta x^* \\
\Delta x & \Delta x^* & \Delta x  & \Delta x^* \\
\Delta x & \Delta x^* & \textbf{j}\Delta x & \textbf{j}\Delta x^* \\
\Delta x & \Delta x^* & \Delta x & \Delta x^* 
\end{array}\right]
\end{equation*}

\noindent It can be observed that the first and the third row of $\Delta \textbf{S}$ are linearly dependent and hence CIDSTC in Example \ref{exple1} is not fully diverse.
\end{example}

\indent From the results of the Proposition \ref{2_d_non_full_diversity}, full diversity of CIDSTC can be obtained by making the real variables $x_{iI}, x_{iQ}$ for $i = 1, 2, \cdots T$ take values from an appropriate multi-dimensional signal set. In particular, the signal set needs to be chosen such that such that $\Delta \textbf{S}$ is full rank for every pair of codewords. The determinant of $\Delta \textbf{S}$ will be a polynomial in variables $\Delta x_{iI}, \Delta x_{iQ}$ for $i = 1, 2, \cdots T$. Therefore, a signal set has to be chosen to make determinant of $\Delta \textbf{S}$ non-zero for every pair of codewords. A particular choice of the signal set depends on the design in use. However, it is to be noted that, more than one multi dimensional signal set can provide full diversity for a given design.

\indent In the rest of this section, we provide a multi-dimensional signal set, $\Lambda$ for the CIDSTC, $\mathcal{C}$ in Example \ref{exple1}  such that, when the variables $x_{1I}, x_{1Q}, \cdots x_{4I}, x_{4Q}$ take values from $\Lambda$, the CIDSTC is fully diverse. Towards that direction, real and imaginary components of \mbox{det}$\left[\Delta \textbf{S}\right]$ for any pair of codewords is given in \eqref{realdet} and \eqref{imagdet} respectively. 

\begin{figure*}

\begin{equation}
\label{realdet}
\mbox{Re }\mbox{det} \left[\Delta \textbf{S}\right] = 4\left(\Delta x_{1I}\Delta x_{3Q} - \Delta x_{1Q}\Delta x_{3I}\right)^{2} - 4\left(\Delta x_{2Q}\Delta x_{4I} - \Delta x_{2I}\Delta x_{4Q}\right)^{2} \mbox{ and }
\end{equation}
\begin{eqnarray}
\label{imagdet}
 \mbox{Im } \mbox{det} \left[\Delta \textbf{S}\right] =  -4\left(\Delta x_{1I}\Delta x_{4Q} - \Delta x_{1Q}\Delta x_{4I}\right)^{2} - 4\left(\Delta x_{2I}\Delta x_{3Q} - \Delta x_{3I}\Delta x_{2Q}\right)^{2} \nonumber \\ + 
8\left(\Delta x_{1I}\Delta x_{2Q} - \Delta x_{1Q}\Delta x_{2I}\right)\left(\Delta x_{3I}\Delta x_{4Q} - \Delta x_{3Q}\Delta x_{4I}\right).
\end{eqnarray}

\end{figure*}

\indent Full diversity for the code in Example \ref{exple1} can be obtained by using a signal set which is carved out of a rotated $\mathbb{Z}^{8}$ lattice such that either the real or the imaginary components of \mbox{det}$\left[\Delta \textbf{S}\right]$ is non zero for any pair of codewords \cite{Vit}. In general, the variable $z_{i}$'s of the vector $\textbf{z} = \left[z_{1}, z_{2} \cdots z_{8} \right]^{T} \in \mathbb{Z}^{8}$ can take values from $\Xi$ say, a M - PAM set where M is any natural number. The $8$-dimensional real vector $\textbf{z}$ is rotated using the generator, $\textbf{G}$ of a rotated $\mathbb{Z}^{8}$ lattice to generate a lattice point $\textbf{l} = \left[ x_{1I} ~ x_{1Q} \cdots ~x_{4I} ~ x_{4Q}\right]$ as $\textbf{l} = \textbf{G}\textbf{z}$ using which a complex vector $\left[x_{1}~x_{2}~ x_{3}~ x_{4}\right]$ is transmitted to all the relays. The signal set $\Lambda$ is identified using computer search as $\left(\textbf{G}, ~\Xi \right)$. As an example, $z_{i}$'s is allowed to take values from $\Xi = \left\lbrace 1, -1\right\rbrace$ and the generator of the lattice, $\textbf{G}$ is found to be in \eqref{latt_gen_1}.
\begin{figure*}
{\small
\begin{equation}
\label{latt_gen_1}
\textbf{G} = \left[\begin{array}{rrrrrrrr}
-0.4081  &  0.4726  &  0.1809  & -0.3955  & -0.1556  & -0.2860  & -0.2408  &  0.5070\\
-0.3256  & -0.0526  & -0.6611  &  0.3368  & -0.5730 &  -0.0934  & -0.0510  &  0.0329\\
-0.3481  &  0.0745  &  0.1031  &  0.3303  &  0.2050  &  0.0771  &  0.7672  &  0.3421\\
-0.3844  & -0.0969  & -0.2606  & -0.6933 &   0.0365 &   0.2074 &   0.3121 &  -0.3905\\
-0.3905  & -0.1319  &  0.6135  &  0.1962  & -0.3501  & -0.2271  &  0.0040  & -0.4909\\
-0.3832  & -0.0759  &  0.1271  &  0.1629  &  0.0892 &   0.7841  & -0.4096  &  0.1193\\
-0.2931  &  0.4104  & -0.2385  &  0.2690  &  0.5999  & -0.2462  & -0.2212  & -0.3834\\
-0.2710  & -0.7532  & -0.0468 &  -0.0540 &   0.3372 &  -0.3654  & -0.1917 &   0.2649
\end{array}\right].
\end{equation}
}
\hrule
\end{figure*}

The matrix \textbf{G} in \eqref{latt_gen_1} is obtained using computer search. Through simulations, it has been verified that if the vector $\left[ x_{1I} ~ x_{1Q} \cdots ~x_{4I} ~ x_{4Q}\right]$ takes value from the above signal set $\left( \textbf{G}, ~\Xi \right) $, then the determinant of $\Delta \textbf{S}$ is non zero for any pair of codewords of $\mathcal{C}$ and hence $\mathcal{C}$ is fully diverse.
In general, for CIDSTCs of any dimension, appropriate signal sets needs to be designed so as to make the code fully diverse.
\section{simulations}
\label{sec5}

\begin{figure}
\centering
\includegraphics[width=3in]{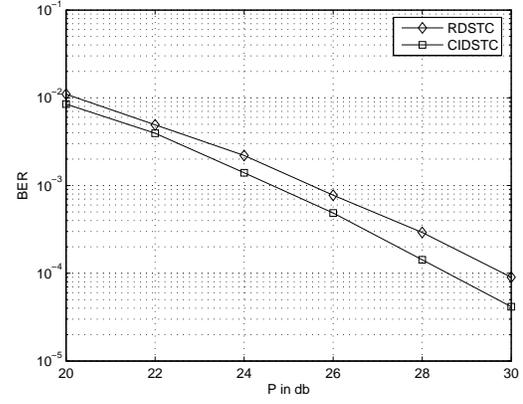}
\caption{BER comparison of CIDSTC with RDSTC for R = 2 and T = 8.}
\label{fig_comp1}
\end{figure}

\begin{figure}
\centering
\includegraphics[width=3in]{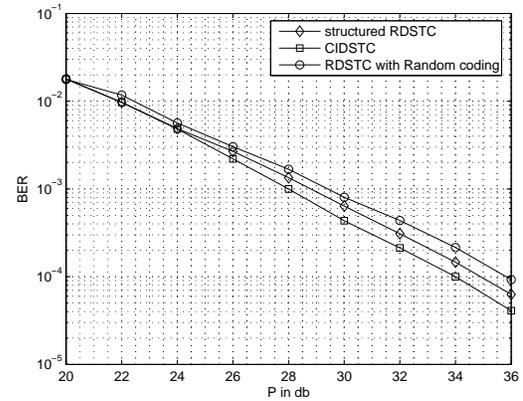}
\caption{BER comparison of CIDSTC with RDSTC for R = 1 and T = 4.} 
\label{fig_comp2}
\end{figure}

In this section, we provide simulation results for the performance comparison of CIDSTC and RDSTC for a wireless network with two relay nodes (Figure \ref{fig_comp1}) and a single relay node (Figure \ref{fig_comp2}). Optimal power allocation strategy discussed in Subsection \ref{subsec3} has been used in our simulation setup though the strategy is not optimal for smaller values of R. Even though the power allocation used is not optimal, CIDSTCs are found to perform better than their corresponding RDSTCs. We have used the Bit Error Rate (BER) which corresponds to errors in decoding every bit as error events of interest. For the network with 2 relays, since we need $T \geq 4R$, for CIDSTC, we use the channel coherence time of $T=8$ channel use for both the schemes.

\indent The real and imaginary parts of information symbols are chosen equiprobably from a 2- PAM signal set $\{-1, 1\}$ and are appropriately scaled to maintain the unit norm condition. Simulations are carried out using the linear designs in variables $x_{1}$, $x_{2} \cdots x_{8}$ as given in \eqref{has_design} and \eqref{CI_design} for RDSTC and CIDSTC respectively. It can be verified that design in \eqref{CI_design} is of the required form given in \eqref{code}.\\
\indent The design in \eqref{has_design} is four group decodable, i.e., the variables can be partitioned into four groups and the ML decoding can be carried out for each group of variables independently of the variables in the groups and the variables of each groups need to be jointly decoded \cite{RaR}. The corresponding four groups of real variables are, $\{x_{1\textit{I}}, x_{4Q}, x_{5\textit{I}}, x_{8Q}\}, \,\{x_{1Q}, x_{4\textit{I}}, x_{5Q}, x_{8\textit{I}}\},\, \{x_{2\textit{I}}, x_{3Q}, x_{6\textit{I}}, x_{7Q}\},\\ \{x_{3\textit{I}}, x_{2Q}, x_{7\textit{I}}, x_{6Q}\}$. We use sphere decoding algorithm for ML decoding \cite{ViB}. Though the design in \eqref{has_design} is four group decodable, the design in \eqref{CI_design} is not four group ML decodable. Full diversity is obtained by making every group of real variables choose values from a rotated $\mathbb{Z}^{4}$ lattice constellation \cite{Vit} whose generator given by,
\begin{equation*}
\textbf{G} = \left[\begin{array}{rrrr}
-0.4316 &  -0.2863  &  0.5857 &  -0.6234\\
-0.6856  & -0.4520  & -0.5445  & 0.1707\\
-0.4479  &  0.8285  & -0.2068  & -0.2647\\
-0.3782  &  0.1649  &  0.5636  &  0.7157\\
\end{array}\right].
\end{equation*}
{\small
\begin{equation}
\label{has_design}
\left[\begin{array}{rrrr}
x_{1} & x_{4}^{*} & x_{5} & x_{8}^{*}\\
x_{2} & x_{3}^{*} & x_{6} & x_{7}^{*}\\
x_{3}^{*} & -x_{2} & x_{7}^{*} &- x_{6}\\
-x_{4}^{*} & x_{1} & -x_{8}^{*} & x_{5}\\
x_{5} & x_{8}^{*} & x_{1} & x_{4}^{*}\\
x_{6} & x_{7}^{*} & x_{2} & x_{3}^{*}\\
x_{7}^{*} & -x_{6} & x_{3}^{*} &- x_{2}\\
-x_{8}^{*} & x_{5} & -x_{4}^{*} & x_{1}\\
\end{array}\right]
\end{equation}
\begin{equation}
\label{CI_design}
\left[\begin{array}{rrrrrrrr}
x_{1} & x_{1}^{*} & x_{4}^{*} & x_{4} & x_{5} & x_{5}^{*} & x_{8}^{*} & x_{8} \\
x_{2} & x_{2}^{*} &  x_{3}^{*} & x_{3} & x_{6} & x_{6}^{*} & x_{7}^{*} & x_{7}\\
x_{3}^{*} & x_{3} & -x_{2} & -x_{2}^{*} & x_{7}^{*} & x_{7} & -x_{6} & -x_{6}^{*}\\
-x_{4}^{*} & -x_{4} &  x_{1} & x_{1}^{*} & -x_{8}^{*} & -x_{8} & x_{5} & x_{5}^{*}\\
x_{5} & x_{5}^{*} & x_{8}^{*} & x_{8} & x_{1} & x_{1}^{*} & x_{4}^{*} & x_{4}\\
x_{6} & x_{6}^{*} &  x_{7}^{*} & x_{7} & x_{2} & x_{2}^{*} & x_{3}^{*} & x_{3}\\
x_{7}^{*} & x_{7} & -x_{6} & -x_{6}^{*} & x_{3}^{*} & x_{3} & -x_{2} & -x_{2}^{*}\\
-x_{8}^{*} & -x_{8} &  x_{5} & x_{5}^{*} & -x_{4}^{*} & -x_{4} & x_{1} & x_{1}^{*}\\
\end{array}\right]\\
\end{equation}
}

\noindent BER comparison of the two schemes using the above designs is shown in Figure \ref{fig_comp1}. The plot shows that CIDSTC in \eqref{CI_design} performs better than the RDSTC in \eqref{has_design} by close to 1.5 to 2 db.\\
\indent Simulation results comparing the BER performance of CIDSTC in Example \ref{exple1} with its corresponding RDSTC is shown in Figure \ref{fig_comp2}. Full diversity is obtained by choosing a rotated $\mathbb{Z}^{8}$ lattice constellation (Section \ref{sec4}) whose generator is given by \eqref{latt_gen_1}. The plot shows the superiority of the design in Example \ref{exple1} over its RDSTC counterpart by 1 db. Simulation results comparing the BER performance of CIDSTC in Example \ref{exple1} with RDSTC from random coding is also shown in Figure \ref{fig_comp2} which shows the superiority of CIDSTC by 1.75 - 2 db at larger values of P.

\section{Discussion}
\label{sec6}
The technique of co-ordinate interleaved distributed space-time coding at the relays was introduced for wireless relay networks having $R$ relays each having two antennas. For $T \geq 4R$, we have shown that CIDSTC provides coding gain compared to the scheme when transmit and receive signals at different antennas of the same relay are processed independently. This improvement is at the cost of only a marginal additional complexity in processing at the relays. Condition on the full diversity of CIDSTCs is also presented. Some of the possible directions for future work is to 
extended the above technique to relay networks where the source and the destination nodes have multiple antennas. 
Also, if the relays have more than two antennas then a general linear processing need to be employed in the place of CIVP and new performance bounds need to be derived.
\appendices

\section{Proof of Lemma \ref{thm2}}\label{append2}

The channel \textbf{h} as given in \eqref{channel} can be written as the product $\textbf{G}\textbf{k}$ of  $\textbf{G}$ and $\textbf{k}$ where
\begin{equation*}
\textbf{G} = \mbox{diag}\left\{ g_{11}, g_{11}, g_{21}, g_{21}, g_{12}, \ldots ,g_{1R}, g_{1R}, g_{2R}, g_{2R}\right\}
\end{equation*}
and
\begin{equation*}
\textbf{k} = \left[ k_{11}, k_{21}, k_{11},- k_{21}, k_{12} \ldots , k_{1R}, k_{2R}, k_{1R},- k_{2R}\right]^T.
\end{equation*}
Since $\textbf{U}$ is Hermitian and positive definite, we can write $\textbf{U} = \textbf{V}^{H}\textbf{D}\textbf{V}$, where $\textbf{D}$ is diagonal matrix containing the eigen values of $\textbf{U},$ in \eqref{chernoff}. Since, $\textbf{U}$ is of full rank, the right hand side of the resulting following PEP expression
\begin{center}
$P\left( \textbf{S}\rightarrow \textbf{S}^\prime\right | g_{ij})\, \leq \,E \,e^{-(P^\prime \textbf{k}^{H}\textbf{G}^{H}\textbf{V}^{H}\textbf{D}\textbf{V}\textbf{G}\textbf{k})}.$ \\
\end{center}
can be upper-bounded by replacing $\textbf{D}$ by $\rho^{2}\textbf{I}_{4R},$  where $\rho^2$ is the minimum singular value of $\textbf{U}.$ Then, we have, 
\begin{center}
$P^\prime  \rho^{2}\textbf{k}^{H}\textbf{G}^{H}\textbf{G}\textbf{k} = \sum_{j = 1}^{R} \left[ P^\prime \rho^{2}g_{j}(|k_{1j}|^{2} + |k_{2j}|^{2})\right] $
\end{center}

\noindent where $g_{j} = |g_{1j}|^{2} + |g_{2j}|^{2}.$\\

\noindent Since the set of random variables $|k_{ij}|^{2}$ are independent (from Proposition \ref{prop}) and distributed as  $2e^{-2|k_{ij}|^{2}},$ we have,
\begin{center}
$P\left( \textbf{S}\rightarrow \textbf{S}^\prime\right | g_{ij})\, \leq  \prod_{j = 1}^{R}\prod_{i = 1}^{2} E \, \left[ e^{-P^\prime \rho^{2}g_{j}(|k_{ij}|^{2})}\right],$ 
\end{center}
and hence
$$P\left( \textbf{S}\rightarrow \textbf{S}^\prime\right | g_{ij})\, \leq  \prod_{j = 1}^{R}\prod_{i = 1}^{2} \int_{0}^{\infty} 2e^{-(2+ P^\prime \rho^{2}g_{j})|k_{ij}|^{2}}d|k_{ij}|^{2}$$
leading to
\begin{center}
$P\left( \textbf{S}\rightarrow \textbf{S}^\prime\right | g_{ij})  \leq  \prod_{j = 1}^{R} 4\left[ \frac{1}{2 + P^\prime \rho^{2}g_{j}}\right] ^{2}.$
\end{center}
This completes the proof.


\section{Proof of Theorem \ref{thm3}}\label{append3}
From \eqref{theorem4eqn} we have
\begin{center}
$P\left( \textbf{S}\rightarrow \textbf{S}^\prime\right | g_{ij})  \leq  \prod_{j = 1}^{R} 4(\frac{1}{2 + P^\prime \rho^{2}g_{j}})^{2}$\\
\end{center}
\noindent where $g_{j} = |g_{1j}|^{2} + |g_{2j}|^{2}.$

\noindent Since $|g_{ij}|^{2}$ are exponentially distributed independent random variables, the random variable $g_{j} =  |g_{1j}|^{2} + |g_{2j}|^{2}$ has the Gamma distribution, $p(g_{j}) = g_{j}e^{-g_{j}}.$


\noindent Since $g_{j}$ are independent, we omit the subscript $j$ and from \eqref{theorem4eqn} we get\\
\begin{center}
$P\left( \textbf{S}\rightarrow \textbf{S}^\prime\right) \, \leq    \left[ 4E\, (\frac{1}{2 + P^\prime \rho^{2}g})^{2}\right] ^{R}$\\
\end{center}
\begin{equation}\label{eqna}
P\left( \textbf{S}\rightarrow \textbf{S}^\prime\right) \, \leq  \left[4 \int_{0}^{\infty} \left( \frac{1}{2 + P^\prime \rho^{2}g}\right) ^{2}ge^{-g}dg\right] ^{R}.
\end{equation}
Let  $\alpha = P^\prime \rho^{2}.$ By change of variables in the integral in \eqref{eqna} as $t = 2 + \alpha g$
we have
\begin{equation*}
P\left( \textbf{S}\rightarrow \textbf{S}^\prime\right) \, \leq  \left[\frac{4}{\alpha^{2}}e^{\frac{2}{\alpha}}\right] ^{R}\left[ \int_{2}^{\infty} \frac{1}{t}e^{-\frac{t}{\alpha}}dt - \int_{2}^{\infty} \frac{1}{t^{2}}e^{-\frac{t}{\alpha}}dt \right]^{R}\\
\end{equation*}
and further, changing $-\frac{t}{\alpha}\:$ to $\:  t, $
\begin{equation}
\label{fullone}
P\left( \textbf{S}\rightarrow \textbf{S}^\prime\right) \, \leq \left[\frac{4}{\alpha^{2}}e^{\frac{2}{\alpha}}\right]^{R} \left[  \int_{-\frac{2}{\alpha}}^{-\infty}\frac{1}{t}e^{t}dt + \frac{2}{\alpha}\int_{-\frac{2}{\alpha}}^{-\infty}\frac{1}{t^{2}}e^{t} dt \right]^{R}.\\
\end{equation}
Using the chain rule of integration, for any integer $m$ we can write the recursive relation,

\begin{equation*}
 \int_{-\frac{2}{\alpha}}^{-\infty}\frac{1}{t^{m + 1}}e^{t}dt  = \frac{1}{m}\left[\int_{-\frac{2}{\alpha}}^{-\infty}\frac{1}{t^m}e^{t}dt + \left( \frac{\alpha}{2}\right) ^{m}\left(-1 \right)^{m}  \right].
\end{equation*}

\begin{equation}
\label{recursive}
 \int_{-\frac{2}{\alpha}}^{-\infty}\frac{1}{t^{2}}e^{t}dt  = \left[\int_{-\frac{2}{\alpha}}^{-\infty}\frac{1}{t}e^{t}dt - \frac{\alpha}{2}\right].
\end{equation}
Applying \eqref{recursive} in \eqref{fullone}, we have

{\small
\begin{equation*}
P\left( \textbf{S}\rightarrow \textbf{S}^\prime\right) \, \leq \left[\frac{4}{\alpha^{2}}e^{\frac{2}{\alpha}}\right]^{R} \left[  \int_{-\frac{2}{\alpha}}^{-\infty}\frac{1}{t}e^{t}dt + \frac{2}{\alpha}\left[\int_{-\frac{2}{\alpha}}^{-\infty}\frac{1}{t}e^{t}dt - \frac{\alpha}{2}\right] \right]^{R}.
\end{equation*}
}

\begin{equation}\label{eqnb}
\leq \left[\frac{4}{\alpha^{2}}e^{\frac{2}{\alpha}}\right]^{R} \left[  \int_{-\frac{2}{\alpha}}^{-\infty}\frac{1}{t}e^{t}dt + \frac{2}{\alpha}\int_{-\frac{2}{\alpha}}^{-\infty}\frac{1}{t}e^{t}dt - 1\right]^{R}.
\end{equation}
We know that 
\begin{equation*}
-Ei( -\frac{2}{\alpha} ) = \int_{-\frac{2}{\alpha}}^{-\infty}\frac{1}{t}e^{t}dt \\
\end{equation*}
is the exponential integral function \cite{GrR}, \cite{JiH1}.\\

As in \cite{JiH1}, for large $P$, $-Ei( -\frac{2}{\alpha} ) = -Ei( -\frac{64R}{PT\rho^{2}}) =  \mbox{log}(P) + O(1) \simeq \mbox{log}(P)$ and  $e^{\frac{2}{\alpha}} \simeq 1.$
Hence, \eqref{eqnb} can be written as
\begin{equation*}
P\left(\textbf{S}\rightarrow \textbf{S}^\prime\right) \, \leq  \left[\frac{64R}{PT\rho^{2}}\right] ^{2R}\left[ \mbox{log}(P) + \frac{64R}{PT\rho^{2}}\mbox{log}(P) - 1\right] ^{R}
\end{equation*}

\begin{equation}\label{eqnc}
 \leq  \left[\frac{64R}{PT\rho^{2}}\right] ^{2R}(\mbox{log}(P))^{R}\left[ 1 + \frac{64R}{PT\rho^{2}}\right]^{R}.\\
\end{equation}

\noindent Considering the most significant term of $P$ in \eqref{eqnc},
\begin{equation}
P\left( \textbf{S}\rightarrow \textbf{S}^\prime\right) \leq \left[\frac{64R}{T\rho^{2}}\right]^{2R}\left[\frac{(\mbox{log}(P))^{R}}{P^{2R}}\right]. 
\end{equation}
This completes the proof.



\begin{IEEEbiography}[{\includegraphics[width=1in,height=1.25in,clip,keepaspectratio]{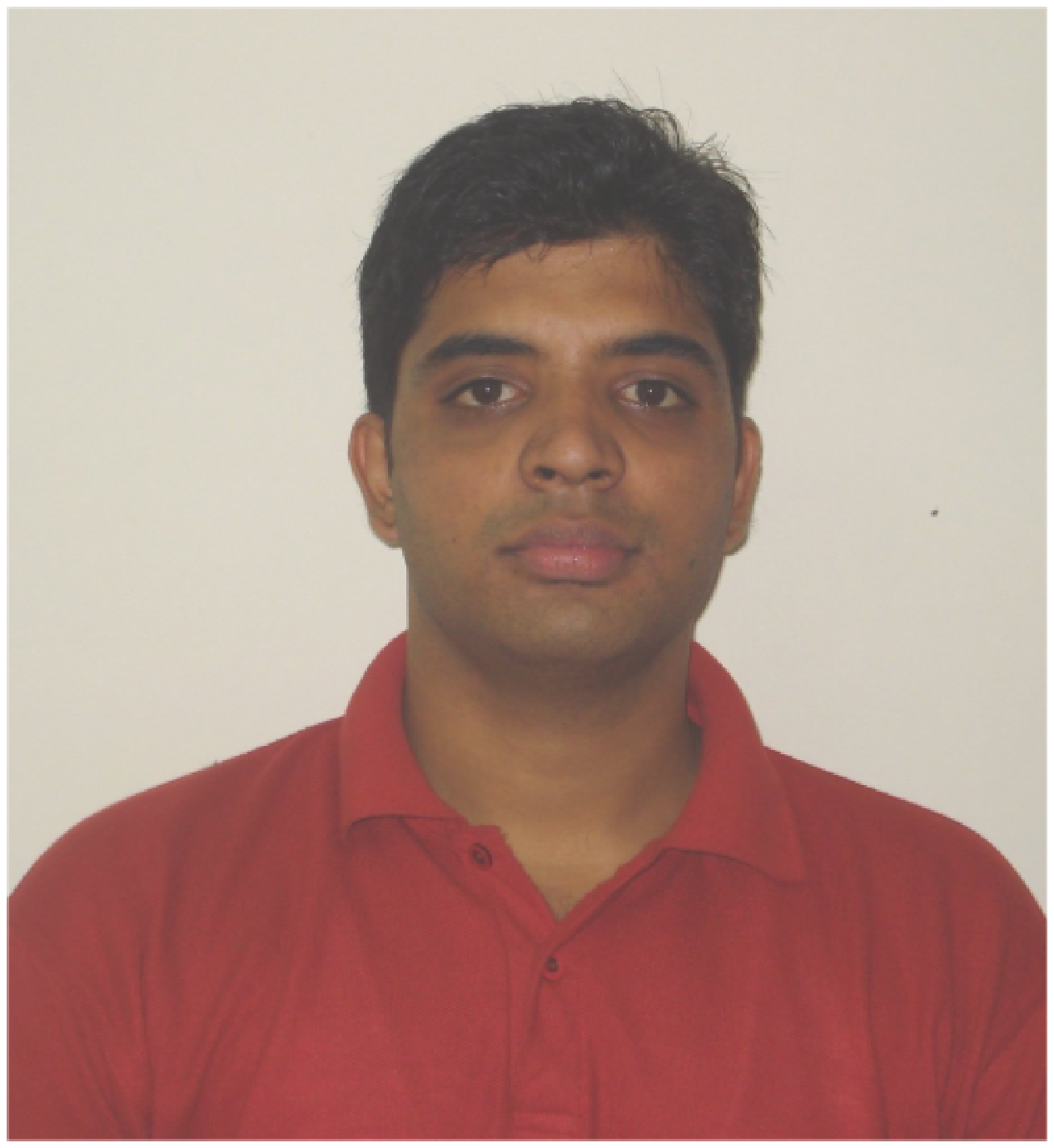}}]{Harshan J}
was born in Karnataka, India. He received the B.E. degree from Visvesvaraya Technological University, Karnataka in 2004. He was working with Robert Bosch (India) Ltd, India till December 2005. He is currently a Ph.D. student in the Department of Electrical Communication Engineering, Indian Institute of Science, Bangalore, India. His research interests include wireless communication, information theory, space-time coding and coding for multiple access channels and relay channels.\\
\end{IEEEbiography}

\begin{IEEEbiography}[{\includegraphics[width=1in,height=1.25in,clip,keepaspectratio]{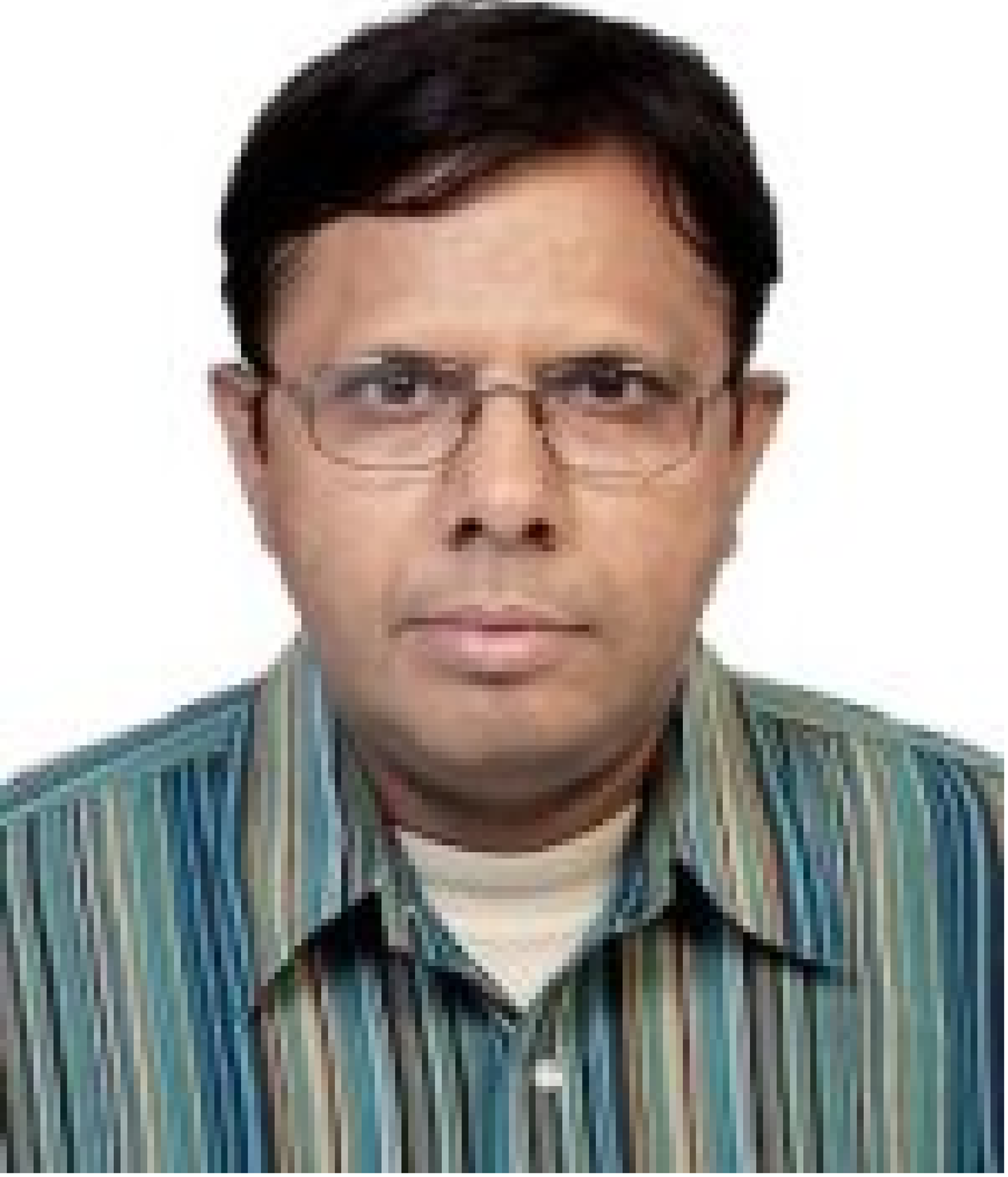}}]{B. Sundar Rajan}
(S'84-M'91-SM'98) was born in Tamil Nadu, India. He received the B.Sc. degree in mathematics from Madras University, Madras, India, the B.Tech degree in electronics from Madras Institute of Technology, Madras, and the M.Tech and Ph.D. degrees in electrical engineering from the Indian Institute of Technology, Kanpur, India, in 1979, 1982, 1984, and 1989 respectively. He was a faculty member with the Department of Electrical Engineering at the Indian Institute of Technology in Delhi, India, from 1990 to 1997. Since 1998, he has been a Professor in the Department of Electrical Communication Engineering at the Indian Institute of Science, Bangalore, India. His primary research interests include space-time coding for MIMO channels, distributed space-time coding and cooperative communication, coding for multiple-access and relay channels, with emphasis on algebraic techniques.

Dr. Rajan is an Associate Editor of the IEEE Transactions on Information Theory, an Editor of the IEEE Transactions on Wireless Communications, and an Editorial Board Member of International Journal of Information and Coding Theory. He served as Technical Program Co-Chair of the IEEE Information Theory Workshop (ITW'02), held in Bangalore, in 2002. He is a Fellow of Indian National Academy of Engineering and recipient of the IETE Pune Center's S.V.C Aiya Award for Telecom Education in 2004. Also, Dr. Rajan is a Member of the American Mathematical Society. 
\end{IEEEbiography}
\vfill

\end{document}